\documentclass[10pt,conference,twocolumn]{IEEEtran}
\usepackage{amsfonts,epsfig,amsmath,latexsym,amssymb,amscd,multirow,graphicx,geometry,lscape,amsmath,amssymb,bm,pifont,graphicx,amssymb,amsmath}
\usepackage{cancel,algorithm,algorithmic,algorithm,verbatim,array,multicol,palatino,amsfonts}
\usepackage{enumitem}
\usepackage[usenames,dvipsnames]{xcolor}
\usepackage{graphicx}
\usepackage{lineno}
\usepackage{subfig}
\usepackage[normalem]{ulem}
\usepackage{graphicx}
\usepackage{mathrsfs}
\usepackage{dsfont}

\newcommand{\x}{\mathbf{x}}

\newcommand{\f}{\mathbf{f}}

\newcommand{\Y}{\mathbf{Y}}

\newcommand{\exE}{\mathbb{E}}

\newcommand{\PSI}{\bm{\Psi}}

\newcommand{\YN}{\Y_{\text{\tiny{1:N} } } }

\newcommand{\fN}{\f_{\text{\tiny{1:N} } } }

\newcommand{\sigW}{\sigma^2_\text{\tiny{W}}}
\newcommand{\sigV}{\sigma^2_\text{\tiny{V}}}

\newcommand{\B}{\mathbf{B}}

\newcommand{\GP}{\mathcal{GP}}

\newfont{\fsc}{eusm10}                         

\newtheorem{proposition}{\noindent \textbf{Proposition}}

\newtheorem{remark}{Remark}

\newtheorem{definition}{\noindent \textbf{Definition}}

\modulolinenumbers[1]

\begin{document}
\title{Multimodal Data Fusion of Non-Gaussian Spatial Fields in Sensor Networks}
\author{\IEEEauthorblockN{Pengfei Zhang$^{1}$, Gareth W.~Peters$^{2}$, Ido Nevat$^{3}$, Keng Boon Teo$^{1}$, Yixin Wang$^{1}$}
1. Institute for Infocomm Research (I2R), Singapore \\
2. Department of Actuarial Mathematics and Statistics, Heriot-Watt University, Edinburgh, UK.\\
3. TUMCREATE, Singapore\\
}
\maketitle
\begin{abstract}
\noindent
We develop a robust data fusion algorithm for field reconstruction of multiple physical phenomena. The contribution of this paper is twofold:
First, we demonstrate how multi-spatial fields which can have any marginal distributions and exhibit complex dependence structures can be constructed.
To this end we develop a model where a latent process of these physical phenomena is modelled as Multiple Gaussian Process (MGP), and the dependence structure between these phenomena is captured through a Copula process. This model has the advantage of allowing one to choose any marginal distributions for the physical phenomenon. Second, we develop an efficient and robust linear estimation algorithm to predict the mean behaviour of the physical phenomena using \textit{rank correlation} instead of the conventional \textit{linear Pearson correlation}. Our approach has the advantage of avoiding the need to derive intractable predictive posterior distribution and also has a tractable solution for the rank correlation values. We show that our model outperforms the model which uses the conventional linear Pearson correlation metric in terms of the prediction mean-squared-errors (MSE). This provides the motivation for using our models for multimodal data fusion.
\newline

\end{abstract}
\section{Introduction}

The term "Internet-of-Things" (IoT) describes several technologies and research disciplines in which the Internet extends into the physical world \cite{chui2010internet, gubbi2013internet}.
 IoT networks consist of sensors that can collect different types of data modalities from the environment. For example, sensors can measure temperature, humidity or pollution particles from environment at same time. Therefore, it has been increasingly important problem to study multimodal sensor networks where different modalities exhibit different statistical distributions. In addition, the correlation between different data can also be taken into account in order to make more accurate inference. However, both of these two tasks are difficult and challenging.

Many works have been developed to understand the dependence of multimodal data in sensor networks. Classical methods include the linear dependence structure between different fields, resulting in linear correlated output, namely multiple output Gaussian Process (GP) \cite{osborne2012real, rasmussen2005gaussian}. However, these classical methods suffer from two main drawbacks that make it infeasible to solve real world challenging problems:
\begin{enumerate}
\item The marginal distribution of GP is Gaussian. However, in many practical cases the Normality assumption is violated. For example, wind field is typically modelled as Weibull distribution \cite{conradsen1984review}, and a Poisson distribution is widely used to model discrete counts of data, e.g., the number of pollution particles in the field \cite{yi2015survey}.
\item GP models only capture the linear Pearson correlation dependence, and do not allow for more complex dependence structures. However, more complicated nonlinear dependence structures might exist in real physical contexts. For example, the extreme pressure in spatial regions and extreme rainfalls cannot be captured using linear dependence structures \cite{kao2008trivariate}.
\end{enumerate}

It is therefore necessary to develop new models to incorporate both the non-Gaussian marginals as well as non-linear dependence structure of multimodal fields. Developing such a model is the main focus of this paper. A general framework of modeling dependencies is to use Copula functions \cite{nelsen2007introduction}. Copula models have become popular because of their ability to separate the marginal distribution from the dependance structure of multivariate distributions. This allows nonlinear dependence structures to be captured and modelled. We develop a hierarchical model where the MGP is used as a latent process while the marginal distribution can be any process, and the dependence between these processes is captured via Copula. We study bivariate processes in this paper, however, it can be easily extended to multiple processes.





\section{Background Features of the Model Formulation} \label{SystemModel}
In this section we present important definitions of some key components used in the model construction, namely related to non-parametric Gaussian Processes, linear dependent Gaussian Processes and parametric Copula models. We also discuss some properties of rank correlations related to Copula processes. These definitions are essential for our wireless sensor network models and our problems and solutions as well.
\begin{definition} [Gaussian Process \cite{rasmussen2005gaussian}]
 \label{definition 1}
 \textsl{
 A Gaussian process is a collection of random variables, any finite number of which have a joint Gaussian Distribution.}
\end{definition}

A Gaussian process is completely specified by its mean function, $\mu\left(x\right)$, and covariance function $k\left(x,x'\right)$ and denoted by
\begin{align*}
 f\left(x\right) \sim \GP \left(\mu\left(x\right),k\left(x,x'\right)\right).
\end{align*}

\begin{definition}  [Linearly Dependent Gaussian Processes \cite{bo2010twin}]
 \label{definition_2}
  \textsl{Given two Gaussian processes $f_1$ and $f_2$, if the correlation between $f_1\left(x\right), \forall x \text{ in the domain of }f_1 $ and $f_2\left(x'\right), \forall x' \text{ in the domain of }f_2$ is :
  \small
  \begin{align*}
  k\left(f_1\left(x\right),f_2\left(x'\right)\right)= \mathbb{E}\left[\left(f_1\left(x\right)-\mu_1\left(x\right)\right)\left(f_2\left(x'\right)- \mu_2\left(x'\right)\right)\right],
  \end{align*}
  \normalsize
  then the two random processes are said to be linearly dependent.
 The dependency structure of the two dependent Gaussian processes is captured via a kernel matrix $K$:
\begin{align*}
   K= \left[
			\begin{array}{cc}
			K_1  & K_{12}\\
			K_{21} & K_2
		  \end{array}
	    \right],
\end{align*}
where $K_1$ and $K_2$ are the correlation matrices within process $1$ and process $2$ respectively; $K_{12}$ and $K_{21}$ are the correlation matrices which capture the cross dependency between process $1$ and process $2$.}
\end{definition}

Furthermore, it will be useful to define a Copula distribution for a multivariate random vector as it provides a means to study dependence structures which are scale-free measures of dependence or concordance, see discussions in \cite{taylor2007multivariate}. In general the term Copula is a Latin noun that means ``a link, tie, bond'' which in the context in which we consider it in this work, is used to link marginal distributions to form a joint dependent distribution model.

\begin{definition}[Copula Distribution]
 \textsl{
A function $C:[0,1]\times \cdots \times[0,1] \mapsto [0,1]$ is a Copula if it satisfies:
\begin{itemize}
\item C is grounded;
\item for ever $i \in \left\{1,\ldots,n\right\}$ and any $u_i \in [0,1]$ one has
\begin{equation*}
C(1,\ldots,1,u_i,1,\ldots,1) = u_i
\end{equation*}
i.e. the marginals are uniform.
\item C is n-increasing, such that for all $(x_1,\ldots,x_n),(y_1,\ldots,y_n) \in [0,1]^n$ with $x_i \leq y_i$ one has
\begin{equation*}
\sum_{i_1=1}^2 \cdots \sum_{i_n=1}^2 (-1)^{i_1+\cdots+i_n}C(u_{1i_1},\ldots,u_{ni_n}) \geq 0
\end{equation*}
where $u_{j1}= x_j$ and $u_{j2} = y_j$ for all $j \in \left\{1,\ldots,n\right\}$.
\end{itemize}
}
\end{definition}

We note that in a bivariate context for instance the notion of groundedness is defined as follows.

\begin{definition}[Grounded Function]
 \textsl{
Consider $S_1$ and $S_2$ as non-empty subsets of $[-\infty,\infty]$. Suppose that $S_i$ has at least element $a_i$, for $i \in \left\{1,2\right\}$. Then a function $G:S_1\times S_2 \mapsto \mathbb{R}$ is grounded if
\begin{equation*}
G(x,a_2)= 0 = G(a_1,y), \;\; \forall (x,y) \in S_1 \times S_2
\end{equation*}
}
\end{definition}

Furthermore, one can also state well known related results as follows for combinations of strictly increasing and decreasing functions, see \cite[2nd Edition, Theorem 2.4.4]{nelsen2007introduction}.

\begin{proposition}[Influence of Increasing and Decreasing Transformations of the Marginals]\label{Prop:Transforms}
 \textsl{
Consider two continuous random variables $X_1$ and $X_2$ with joint Copula given by $C_{X_1,X_2}$. If $T_1(\cdot)$ and $T_2(\cdot)$ are two strictly monotone functions defined on Ran$X_1$ and Ran$X_2$, respectively. Then $C_{T_1(X_1),T_1(X_1)}$ is characterized by one of the following combinations:
\begin{itemize}
\item If $T_1$ is strictly increasing and $T_2$ is strictly decreasing, then
\begin{equation*}
C_{T_1(X_1),T_2(X_2)}\left(u_1,u_2\right) = u_1 - C_{X_1,X_2}\left(u_1,1-u_2\right)
\end{equation*}
\item If $T_1$ is strictly decreasing and $T_2$ is strictly increasing, then
\begin{equation*}
C_{T_1(X_1),T_2(X_2)}\left(u_1,u_2\right) = u_2 - C_{X_1,X_2}\left(1-u_1,u_2\right)
\end{equation*}
\item If $T_1$ and $T_2$ are strictly decreasing, then
\begin{align*}
C_{T_1(X_1),T_2(X_2)}\left(u_1,u_2\right)& = u_1 + u_2 - 1 \\&+ C_{X_1,X_2}\left(1-u_1,1-u_2\right)
\end{align*}
\end{itemize}
}
\end{proposition}
%
%
\begin{remark}
It was shown in \cite{taylor2007multivariate} that all the axioms that a concordance measure (measure of dependence) should satisfy, as outlined by \cite{scarsini1984measures}, are also uniquely characterized by a Copula formulation. This means that all known measures of dependence such as familiar correlations, associations, tail dependence and beyond can be captured uniquely by the Copula function.
\end{remark}

Under Copula, rank correlations have the following properties, such as Spearman correlation.

\begin{proposition}[Spearmann's Rho Rank Correlation Under Monotonic Marginal Transforms]\label{SpearmanCorr}
\textsl{
Consider two continuous random variables $X_1$ and $X_2$ with joint copula given by $C_{X_1,X_2}$ with copula density $c(u_1,u_2)$ when it exists. If $T_1(\cdot)$ and $T_2(\cdot)$ are two strictly monotone functions defined on Ran$X_1$ and Ran$X_2$, respectively. Then the Spearmann's rho rank correlation between $X_1$ and $X_2$, denoted by $\rho^S_{X_1,X_2}$, is given after transformation by:
\begin{itemize}
\item If $T_1$ and $T_2$ are strictly increasing, then
\begin{equation*}
\rho^S_{T_1(X_1),T_2(X_2)} = \rho^S_{X_1,X_2}.
\end{equation*}
\item If $T_1$ is strictly increasing and $T_2$ is strictly decreasing, then
\begin{equation*}
\rho^S_{T_1(X_1),T_2(X_2)} = 3 - 12\int_0^1 \int_0^1 C(u_1,1-u_2)\; du_1 du_2
\end{equation*}
\item If $T_1$ is strictly decreasing and $T_2$ is strictly increasing, then
\begin{equation*}
\rho^S_{T_1(X_1),T_2(X_2)} = 3 - 12\int_0^1 \int_0^1 C(1-u_1,u_2)\; du_1 du_2
\end{equation*}
\item If $T_1$ and $T_2$ are strictly decreasing, then
\begin{equation*}
\rho^S_{T_1(X_1),T_2(X_2)} = 12\int_0^1 \int_0^1 C(1-u_1,1-u_2)\; du_1 du_2 -3
\end{equation*}
\end{itemize}
}
\end{proposition}

\begin{IEEEproof}
The proof of each result follows directly from the application of the identity for Spearman's rho linear correlation written in terms of a copula as denoted in \cite{schweizer1981nonparametric},
\begin{equation}
\begin{split}
\rho^S_{X_1,X_2} &= 12\int_0^1 \int_0^1 C(u_1,u_2)\; du_1 du_2 - 3\\
&= 3 - 6\int_0^1 \int_0^1\left[ u_1\frac{\partial C}{\partial u_1}(u_1,u_2) + u_2\frac{\partial C}{\partial u_2}(u_1,u_2)\right] \\&\times du_1 du_2\\
&= 12 \mathbb{E}[U_1U_2] - 3\\
&= \frac{\mathbb{E}[U_1U_2] - \mathbb{E}[U_1]\mathbb{E}[U_2]}{\sqrt{\textrm{Var}(U_1)\textrm{Var}(U_2)}}
\end{split}
\end{equation}
and then application of Proposition \ref{Prop:Transforms} to obtain for each case enumerated.
\end{IEEEproof}

As with Kendall's tau rank correlation, for many Copula families the explicit solution for the Copula based expression for the Spearman rank correlation is known explicitly in terms of the Copula parameters.

Furthermore, it will be often useful to link the rank correlation such as Spearman's rho to the notion of linear correlation that we will denote generically as $\rho$. In general, where the joint dependence structure of the multivariate distribution is specified in terms of a correlation matrix, such as elliptical families where $\rho$ is a model parameter. Then one obtains $\rho^S$ and $\rho$ as given by the identity:
\begin{equation*}
\rho^S(X_1,X_2) = \rho\left(F_1(X_1),F_2(X_2)\right).
\end{equation*}
In certain cases there is also a direct relationship known between rank and linear correlations such as in the multivariate Gaussian Copula case in which the Spearman correlation $\rho_S$ is obtained in terms of $\rho$ linear correlation according to the expression
\begin{equation}\label{rs2r}
\rho = 2\sin\left(\frac{\pi}{6} \rho^S\right).
\end{equation}

Based on these definitions we can now present our hierarchical model for multimodal spatial fields.

\section{Hierarchical Bayesian Model for Multiple Modality Spatial Random Fields}
The sensor network is deployed in $\mathbb{R}^2$ to monitor various physical phenomena. Based on the observations collected by the sensors, we wish to make predictions about the physical quantities at any location in space, denoted $\x_*\in \mathbb{R}^2$. To make the exposition simple we only consider two physical phenomena, but our model can be generalised to any number of modalities.

\begin{enumerate}[noitemsep]
\item The two physical phenomena of interest, denoted $Z^{\left(1\right)}\left(\x_i\right)$ and $Z^{\left(2\right)}\left(\x_j\right)$, are correlated via two latent dependent GPs, $f^{\left(1\right)}\left(\x_i\right) \text{ and } f^{\left(2\right)}\left(\x_j\right)$, at any point
    $\left(\x_i, \x_j \right)\in \mathbb{R}^2$ through a Copula process which we will specify later.
    The two latent GPs $f^{\left(1\right)}\left(\x_i\right),  f^{\left(2\right)}\left(\x_j\right)$ are coupled as per Definition \ref{definition_2}:
\begin{align*}
\begin{split}
&\left(f^{\left(1\right)}\left(\x_i\right),f^{\left(2\right)}\left(\x_j\right) \right) : \mathbb{R}^2 \times \mathbb{R}^2 \mapsto  \mathbb{R}\times\mathbb{R}
\;
\text{ s.t.}\\
&\left(f^{\left(1\right)}\left(\x_i\right),f^{\left(2\right)}\left(\x_j\right) \right) \\& \sim
\GP\left(
\left[
\begin{array}{c}
		\mu^{\left(1\right)}\left(\x_i\right)\\
		\mu^{\left(2\right)}\left(\x_j\right)
		\end{array}
\right],\ K\left(\x_i,\x_j;\PSI\right)\right),
\end{split}
\end{align*}		
where
$\mu^{\left(1\right)}\left(\x_i\right),\mu^{\left(2\right)}\left(\x_j\right) \in \mathbb{R}$ are the mean functions of each of the two GPs. The spatial dependence between any two points is given by the covariance function $\ K \left(\x_i,\x_j;\PSI\right): \mathbb{R}^{2} \times \mathbb{R}^{2} \mapsto  \mathbb{R}$, parameterised by $\PSI$ \cite{rasmussen2005gaussian} and,
\begin{align*}
K=  \left[
		{\begin{array}{cc}
		K^{\left(1\right)} & K^{\left(1,2\right)}\\
		K^{\left(2,1\right)} & K^{\left(2\right)}\\
		\end{array} }
		\right].
	\end{align*}	
\item The two physical phenomena $Z^{\left(1\right)}\left(\x_i\right)$ and $Z^{\left(2\right)}\left(\x_j\right)$ are associated with $f^{\left(1\right)}\left(\x_i\right)$ and $f^{\left(2\right)}\left(\x_j\right)$ through the following Gaussian Copula processes:\\

 $f^{\left(1\right)}\left(\x_i\right)$ and $f^{\left(2\right)}\left(\x_j\right)$ are mapped to $\left[0, 1\right]$ through univariate normal CDFs. Denote the resulting data as $U^{\left(1\right)}\left(\x_i\right)$ and $U^{\left(2\right)}\left(\x_j\right)$. Then we take inverse CDFs at $U^{\left(1\right)}\left(\x_i\right)$ and $U^{\left(2\right)}\left(\x_j\right)$ and denote the resulting data as $Z^{\left(1\right)}\left(\x_i\right)$ and $Z^{\left(2\right)}\left(\x_j\right)$.\\
To summarize, the model for the data generated has the following two-step process:
\small
\begin{align*}
\begin{split}
&\text{Step 1}:\\
&\left[U^{\left(1\right)}\left(\x_i\right),U^{\left(2\right)}\left(\x_j\right)\right] := \left[F_1\left(f^{\left(1\right)}\left(\x_i\right)\right),F_2 \left(f^{\left(2\right)}\left(\x_j\right)\right)\right].\\
&\text{Step 2}:\\
&\left[Z^{\left(1\right)}\left(\x_i\right),Z^{\left(2\right)}\left(\x_j\right)\right] \\&:= \left[H_{1}^{-1}\left(U^{\left(1\right)}\left(\x_i\right)\right),H_{2}^{-1}\left(U^{\left(2\right)}\left(\x_j\right)\right)\right].\\
 \end{split}
 \end{align*}
 \normalsize
  $F_1 \text{ and } F_2$ are marginal CDFs of $f_i \text{ and } f_j$, and $H_{1}^{-1}\left(u_i\right)$ and $H_{2}^{-1}\left(u_j\right)$ represent some inverse CDFs which may be different, $u_i, u_j \in \left[0, 1\right]$.

  Denote the joint CDF of $Z^{\left(1\right)}\left(\x_i\right),Z^{\left(2\right)}\left(\x_j\right)$ as $H_{12}$ and the joint CDF of $f^{\left(1\right)}\left(\x_i\right),f^{\left(2\right)}\left(\x_j\right)$ as $F_{12}$.\\
  The above is a Copula process and by Sklar's Theorem,
  \begin{align*}
   C^{GA}\left(u_i, u_j\right)
   &= F_{12}\left(F_1^{-1}\left(u_i\right),F_2^{-1}\left(u_j\right)\right).\\
   C^{GA}\left(u_i, u_j \right)
   &= H_{12}\left(H_1^{-1}\left(u_i\right),H_2^{-1}\left(u_j\right)\right).
  \end{align*}
  $ C^{GA}\left(u_i, u_j\right)$ refers to Gaussian Copula.
\item \textbf{Sensors observations:} there are $n_{1}$ sensors measuring the first physical phenomenon and $n_2$ sensors measuring the second physical phenomenon over a $2$-D region $\mathcal{X} \subseteq \mathbb{R}^2$, at locations
$\x_{i} \in \mathcal{X}, i=\left\{1,\cdots, n_1\right\}$ and $\x_{j} \in \mathcal{X}, j=\left\{1,\cdots, n_2\right\}$ , assumed known.
Each sensor collects a noisy observation of the respective physical process:
\begin{align*}
Y^{\left(1\right)}\left(\x_i\right) &= Z^{\left(1\right)}\left(\x_i\right) + W,\\
Y^{\left(2\right)}\left(\x_j\right) &= Z^{\left(2\right)}\left(\x_j\right) + V,
\end{align*}
where $W$ and $V$ are i.i.d Gaussian noises:
 $W\sim N\left(0,{\sigW}\right)$ , $V\sim N\left(0,{\sigV}\right)$.
\item We denote by $\Y$ the observation vector of the two physical phenomena, as follows:
\begin{align*}
\Y =
\left[
   \underbrace{Y_{1}^{\left(1\right)}, Y_{2}^{\left(1\right)}, \ldots,  Y_{n_1}^{\left(1\right)}}_{\text{phenomenon }1},
   \underbrace{Y_{1}^{\left(2\right)}, Y_{2}^{\left(2\right)}, \ldots,  Y_{n_2}^{\left(2\right)}}_{\text{phenomenon }2}
\right]^\top.
\end{align*}
\end{enumerate}

\section{Estimation Objectives} \label{Estimation Objectives}
The goal is to derive a low complexity algorithm to perform multimodal spatial field reconstitution, given noisy observations of the two physical phenomena $\Y$.  the objective is to make predictions for the intensities $f_*^{\left(1\right)}:= f^{\left(1\right)}\left(\x_*\right)$ and $f_*^{\left(2\right)}:= f^{\left(2\right)}\left(\x_*\right)$ of the phenomena at any location $\x_*$ in the field. To obtain this, we define the following estimation objective:
The Minimum Mean Squared Error (MMSE) estimator of the joint predicted values of intensities at any location $\x_*$:
\small
    \begin{align*}
    \begin{split}
    \hat{\f}_{*} &= \mathbb{E}\left[\f_*|\Y,\x,\x_*,\Theta\right]
                             = \int_{-\infty}^{\infty} \f_* p\left(\f_*|\Y,\x,\x_*,\Theta\right)d\f_*,
    \end{split}
    \end{align*}
		\normalsize
We define the following shorthand notations:
\small
\begin{align*}
\x_* &:=\left(\x_*^{\left(1\right)},\x_*^{\left(2\right)}\right) \text{- test locations}.\\
\f_* &:=\left(f_*^{\left(1\right)},f_*^{\left(2\right)}\right) \text{- predictions of the intensities at } \x_*.\\
\mathbf{Z}_* &:=\left(Z_*^{\left(1\right)},Z_*^{\left(2\right)}\right),\text{predictions of the two phenomena at } \x_*.\\
\x &:=\left(\x^{\left(1\right)},\x^{\left(2\right)}\right) \text{- sensor locations}. \\
\f &:= \left(\f_{1:n_1}^{\left(1\right)},\f_{1:n_2}^{\left(2\right)}\right) \\&\text{- realizations of the Gaussian Processes at } \x.
\end{align*}
\normalsize
To derive the above estimation objectives, the joint predictive density $p\left(\f_{*}|\Y,\x,\x_{*},\Theta\right)$ needs to be evaluated first.
\subsection{Predictive posterior density of the spatial intensities}
The predictive posterior density is given by
\small
\begin{align*}
&p\left(\f_{*}|\Y,\x,\x_{*},\Theta\right)
=\int p\left(\f_{*}|\f,\x,\x_{*},\Theta\right)
       p\left(\f|\Y,\x,\x_{*},\Theta\right)d\f\\
&=\int p\left(\f_{*}|\f,\x,\x_{*},\Theta\right)
      \frac{p\left(\Y|\f,\x,\x_{*},\Theta\right)
			p\left(\f|\x,\x_{*},\Theta\right)}
          { \int p\left(\Y|\f,\x,\x_{*},\Theta\right)p\left(\f|\x,\x_{*},\Theta\right)d\f} d\f.
							\end{align*}
   \normalsize
Unfortunately, the predictive posterior density cannot be evaluated analytically as this involves a $\left(n_1+n_2\right)$-dimensional integral that is intractable. Instead, in the following we develop the Spatial Best Linear Unbiased Estimator (S-BLUE), .

%
%
%

\subsection{Spatial Best Linear Unbiased Estimator (S-BLUE) Field Reconstruction Algorithm} \label{S_BLUE}
We develop the spatial field reconstruction via S-BLUE, which enjoys a low computational complexity and is the optimal estimator (in terms of minimising the MSE) out of all linear estimators. The big advantage of the S-BLUE is that it does not require calculating the predictive posterior density, but only the first two cross moments of the model.
The S-BLUE is the optimal (in terms of minimizing Mean Squared Error (MSE)) of all linear estimators and is given by the solution to the following optimization problem:
\begin{equation}
\label{S_BLUE_objective}
\widehat{f}_* :=\widehat{a} + \widehat{\B} \YN = \arg \min_{a, \B} \exE\left[\left(f_*- \left(a + \B \YN\right)\right)^2\right],
\end{equation}
where $\widehat{a} \in \mathbb{R}$ and $\widehat{\B} \in \mathbb{R}^{1 \times N}$.

The optimal linear estimator that solves (\ref{S_BLUE_objective}) is given by
\begin{align}
\begin{split}
\label{s_blue_estimate}
\hat{f_*}&=\exE_{f_*\; \YN}\left[f_* \; \YN\right]\exE_{\YN}\left[ \YN\; \YN\right]^{-1}\left(\YN-\exE\left[\YN\right]\right),
\end{split}
\end{align}
and the Mean Squared Error (MSE) is given by
\begin{align}
\begin{split}
\label{s_blue_estimate_MSE}
\sigma^2_{*}&=k\left(\x_*,\x_*\right)-
\exE_{f_*\; \YN}\left[f_*\; \YN\right]\exE_{\YN}\left[\YN\; \YN \right]^{-1}\\&\times\exE_{\YN\;f_*}\left[\YN \; f_*\right].
\end{split}
\end{align}

To evaluate (\ref{s_blue_estimate}-\ref{s_blue_estimate_MSE}) we need to calculate the cross-correlation $\exE_{f_*, \YN}\left[f_*\; \YN\right]$, auto-correlation $\exE_{\YN}\left[\YN\; \YN^T \right]$ and $\exE\left[\YN\right]$.

Note, here without loss of generality we calculate the correlation for zero-mean Gaussian process ($\mu_f(x_*)=0$). For the case where $\mu_f(x_*)\neq 0$, it is easy to shift the estimation by $\mu_f(x_*)$.

\subsection{Copula Fitting}
We adopt the approach in \cite{cruz2014fundamental} to fit Gaussian Copula on $\Y$. The approach is below:

\begin{enumerate}
\item Estimate the rank correlation, either $\rho_S(Y_i, Y_j)$ or $\rho_K(Y_i, Y_j)$, for each marginal pair of variables. Then transform to the linear correlation measure;
\item Construct the estimated sample pseudo correlation matrix $\hat{R}^*$ with $(i,j)-th$ element given by Eq. (\ref{rs2r}).
\item The pseudo correlation matrix $\hat{R}^*$ must be made positive definite with unit diagonal entries and off-diagonal entries in the range [-1, 1].
\end{enumerate}

After we fit the Copula, we could estimate the length scale $l$ for square exponential kernel $k(\cdot, \cdot)$ which minimize $\sum (\hat{R}^*_{i,j}-\exp((X_i, X_j)/l^2))$.

\subsection{Cross-correlation and auto-correlation derivations}
In this section, we derive the cross correlation and auto correlations of the terms required in (\ref{s_blue_estimate}) and (\ref{s_blue_estimate_MSE}).

\subsubsection{Cross-correlation between a test point and sensors observations $\exE_{\f_*, \YN}\left[\f_*\; \YN\right]$}

it has been shown that Kendall ($\rho_K$) or Spearman ($\rho_S$) correlation are robust approximation of population correlation $\rho$. For the bivariate normal distribution, there is analytic relationship between these variables as shown in Section \ref{SystemModel}. In \cite{xu2010comparison}, it was shown that $\rho_K$ and $\rho_S$ are invariant to impulse noise.

\begin{proposition}[Cross Correlation 1]\label{CrossCorr1}
\textsl{The cross correlation between a test point and sensors observations $\exE_{\f_*, \YN}\left[\f_*\; \YN\right] = \rho^S_{\f_*,\f} * \sigma_{\f_*}* \sigma_{\YN}$
}
\end{proposition}

\begin{IEEEproof}
According to the expectation definition,
$\exE_{\f_*, \YN}\left[\f_*\; \YN\right] = \exE_{\f_*, \YN}\left[\f_*\; H(F^{-1}(\f))\right] $, this quantity is intractable. Another way of expressing the cross correlation is $\exE_{\f_*, \YN}\left[\f_*\; \YN\right]  =  \rho_{\f_*,\YN} * \sigma_{\f_*}* \sigma_{\YN}$
where $\rho_{\f_*,\YN}$ is the population correlation between  a test point $\f_*$ and observation $\YN$.  It is also difficult to get the population correlation. We use the Spearman rank correlation $\rho^s_{\f_*,\YN}$ to approximate this quantity. Also according to \cite{xu2010comparison}, $\rho_K$ and $\rho_S$  and robust approximation to $\rho$ that are invariant to impulse noise. Through this way, the properties of Spearman correlation can be used.

According to Proposition \ref{SpearmanCorr},
If $T_1$ and $T_2$ are strictly increasing, then
\begin{equation}
\rho^S_{T_1(X_1),T_2(X_2)} = \rho^S_{X_1,X_2}.
\end{equation}
By definition, $H(F^{-1}(\f)$ is strictly increasing function on $\f$, therefore $\rho^s_{\f_*,\YN} = \rho^s_{\f_*,H(F^{-1}(\f))}=\rho^s_{\f_*,\f}$.
Therefore, $\exE_{\f_*, \YN}\left[\f_*\; \YN\right]=\rho^S_{\f_*,\f} * \sigma_{\f_*}* \sigma_{\YN}$.
\end{IEEEproof}

\subsubsection{Correlation of sensors observations $\exE_{\YN}\left[\YN \; \YN^T \right]$}

\begin{proposition}[Cross Correlation 2]\label{CrossCorr2}
\textsl{The cross correlation between sensors observations $\exE_{\YN}\left[\YN\; \YN^T\right] = \rho^S_{\f,\f} * \sigma_{\YN}* \sigma_{\YN}$}
\end{proposition}

\begin{IEEEproof}
Similarly as Proposition \ref{CrossCorr1},
\begin{align*}
\exE_{\YN}\left[\YN\; \YN^T\right]  &= \rho^S_{\YN} *  \sigma_{\YN}* \sigma_{\YN}\\
& = \rho^S_{F^{-1}(H(\f)),F^{-1}(H(\f))} *  \sigma_{\YN}* \sigma_{\YN}\\
& = \rho^S_{\f,\f} * \sigma_{\YN}* \sigma_{\YN}.
\end{align*}
\end{IEEEproof}
%
\subsubsection{Expected value of the observations $\exE_{\YN}\left[\YN\right]$}
$\exE_{\YN}\left[\YN\right]$ is based on the distribution of marginals. For example, if $\YN$ has exponential marginal, then $\exE_{\YN}\left[\YN\right] = 1/\lambda$.
If $\YN$ has gamma marginal, then $\exE_{\YN}\left[\YN\right] = \alpha \beta$, where $\alpha$ and $\beta$ are the shape and rate parameters of Gamma distribution.

\subsubsection{MMSE estimate of predicted intensity values}
The MMSE estimate of the predictions at any location $\x_*$ is given by
\begin{align}
\begin{split}
\label{s_robustblue_estimate_MSE}
\sigma^2_{*}&=k\left(\x_*,\x_*\right)-
\exE_{f_*\; \YN}\left[f_*\; \YN\right]\exE_{\YN}\left[\YN\; \YN \right]^{-1}\\&\times\exE_{\YN\;f_*}\left[\YN \; f_*\right].
\end{split}
\end{align}
where all the quantities have been derived in the above subsections.
%
%

\section{Simulation Results}
In this section, we present simulation results to compare the performance between robust BLUE (R-BLUE) performance and linear BLUE (L-BLUE) performance. L-BLUE is developed by approximating $\rho_{\f_*\; \YN}$ in  Proposition \ref{CrossCorr1} and  $\rho_{\YN\;\YN}$ in Proposition \ref{CrossCorr1} with $\rho^P_{\f_*\; \fN}$ and $\rho^P_{\fN\;\fN}$ respectively. $\rho^P$ denotes the linear Pearson-Norman correlation. We using MSE as the performance metrics. The comparison between R-BLUE and L-BLUE for single GP is studied first, followed by comparison between L-BLUE and R-BLUE for bivariate GP fields. Lastly, we also summarize the MSE comparison for many different realisations.

\subsection{Linear-BLUE and Robust-BLUE Comparison}

In this section, we compare the MSE preformance of the Linear BLUE and the Robust BLUE in terms of single GP field reconstruction accuracy. We run $1000$ realisations and the MSE for robust BLUE is $1.8617$ and linear BLUE is $2.0323.$

%
%
%

\begin{figure}
    \centering
        \epsfysize=4.5cm
        \epsfxsize=7cm
        \epsffile{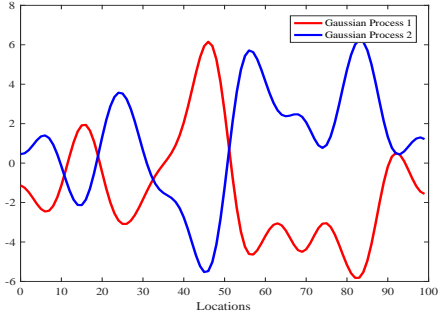}
        \caption{Gaussian process realisations for both modality}
    \label{fig:GPs}
\end{figure}
\begin{figure}
    \centering
        \epsfysize=4.5cm
        \epsfxsize=7cm
        \epsffile{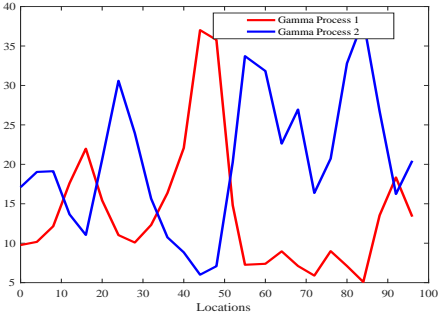}
        \caption{Gamma process realisations for both modality}
    \label{fig:gammas}
\end{figure}
\begin{figure}
    \centering
        \epsfysize=4.5cm
        \epsfxsize=7cm
        \epsffile{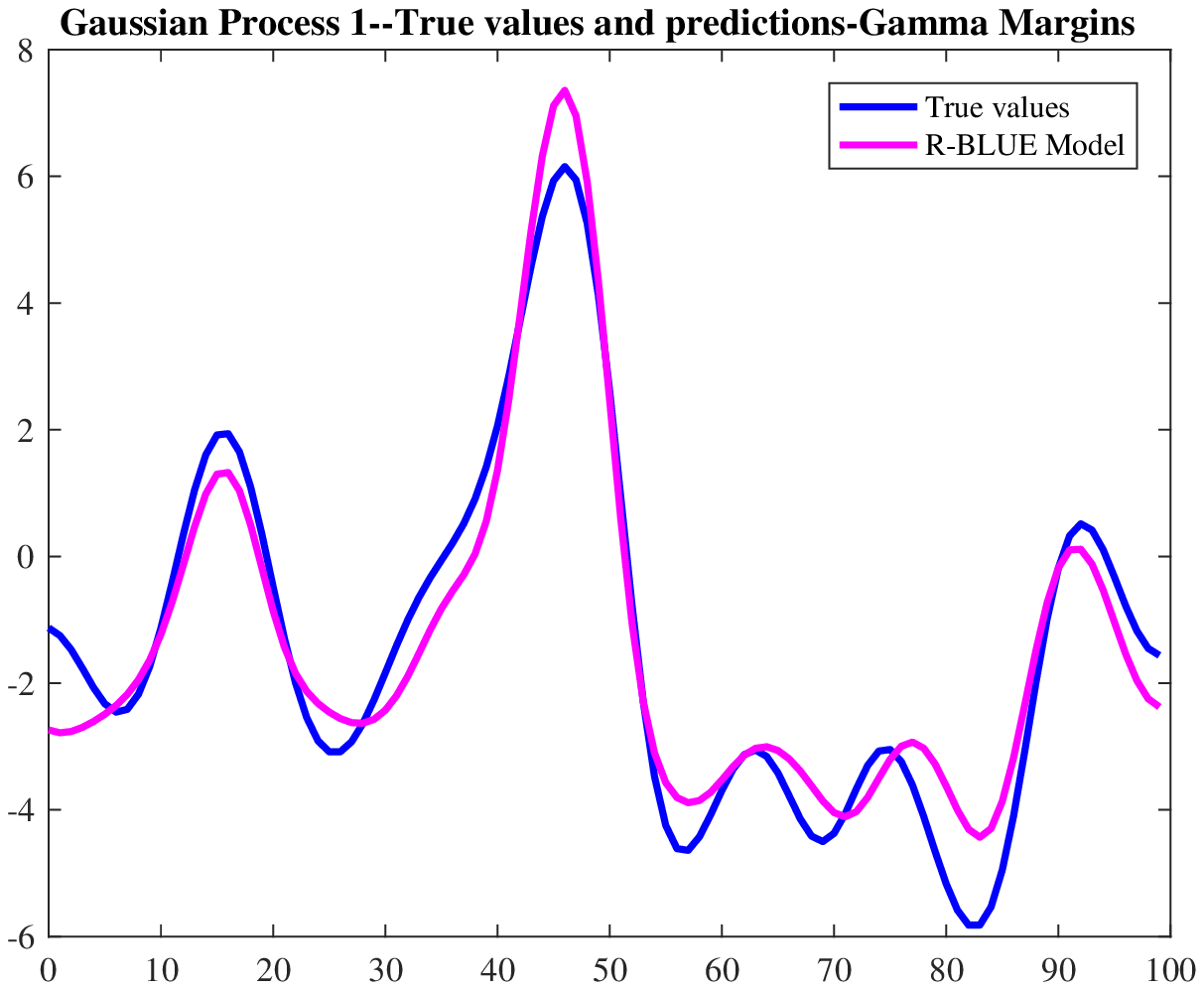}
        \caption{Gaussian process 1 predictions}
    \label{fig:GP1}
\end{figure}
\begin{figure}
    \centering
        \epsfysize=4.5cm
        \epsfxsize=7cm
        \epsffile{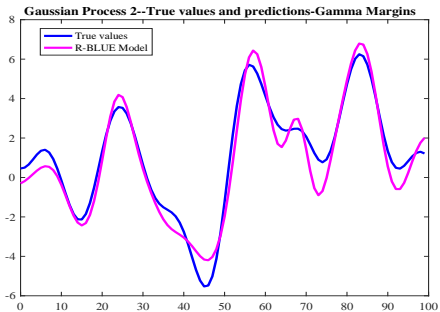}
        \caption{Gaussian process 2 predictions}
    \label{fig:GP2}
\end{figure}

We also run the comparison between Linear BLUE and Robust BLUE for bivariate Gamma process setting.  We generate bivariate GP as shown in Fig. \ref{fig:GPs}, then we transform them into bivariate Gamma processes as shown in Fig. \ref{fig:gammas}. We then reconstruct the GP from the gamma process realisations for two processes as shown in Figs. \ref{fig:GP1} and \ref{fig:GP2}.  After $1000$ iterations, the MSE for GP1 and GP2 using robust BLUE are $1.1711$ and $1.1963$ respectively. The MSE for GP1 and GP2 using linear BLUE are $1.2029$ and $1.2035$ respectively.

We also test the robustness when one of the points is corrupted by impulsive noise. In this case, we purposely distorted a single observation by adding $30$ to its real value. We then ran $1000$ iterations, and the MSE for GP1 and GP2 using robust BLUE are $1.1946$ and $6.2631$ respectively. The MSE for GP1 and GP2 using linear BLUE are  $1.2256$ and $6.8120$ respectively.

\subsection{MSE performance for various parameters under impulsive noise}

\begin{figure}
    \centering
        \epsfysize=4.5cm
        \epsfxsize=7cm
        \epsffile{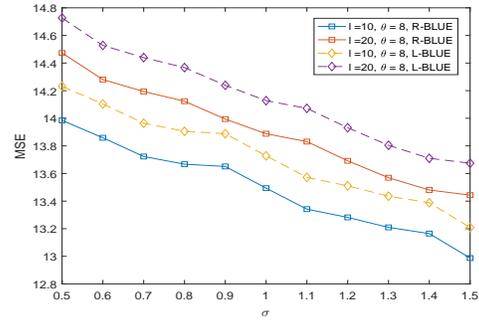}
        \caption{Comparison of MSE with different L and $\sigma$}
    \label{fig:MSEvsL}
\end{figure}

\begin{figure}
    \centering
        \epsfysize=4.5cm
        \epsfxsize=7cm
        \epsffile{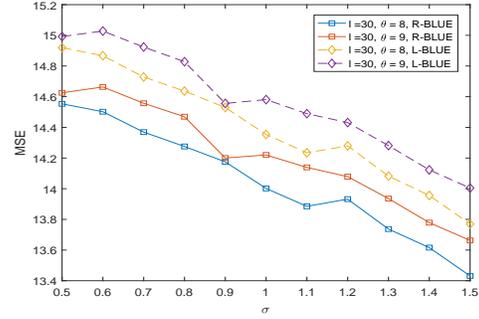}
        \caption{Comparison of MSE with different $\theta$ and $\sigma$}
    \label{fig:MSEvsTheta}
\end{figure}

We also compared the average MSE performance for different sets of parameters, including the length scale ($l$) and scaling factor ($\theta$) as well as the noise $\sigma$. We added impulsive noise equal to amplitude $20$ at location $11$ of signal and showed the robustness of R-BLUE compared with L-BLUE. Both Figs. \ref{fig:MSEvsL} and \ref{fig:MSEvsTheta} show that the R-BLUE provides smaller MSE compared with L-BLUE despite in the presence of impulsive noise.

\section{Conclusions}

We developed efficient data fusion algorithm for field reconstruction in multimodal sensor networks where complex depdeance exists between multimodal fields. W developed low complexity Robust-BLUE method for field reconstruction where dependance is captured through rank correlation. Through extensive simulations, we showed the accuracy of using R-BLUE method and better performance over L-BLUE method which uses traditional Pearson correlation metric in terms of the prediction of mean-squared-errors (MSE).

\bibliographystyle{IEEEtran}
\bibliography{references}
\end{document}